\documentstyle[12pt]{article}

\begin{document}

\begin{titlepage}
{\small
\centerline{\bf CENTRE DE PHYSIQUE THEORIQUE - CNRS - Luminy, Case 907}
\centerline{\bf F-13288 Marseille Cedex 9 - France }
\centerline{\bf Unit\'e Propre de Recherche 7061}
}
\vspace{1cm}

\begin{center}
        {\large \bf The angular momentum operator in the Dirac equation}

\vspace{0.3 cm}

V\'{\i}ctor M. Villalba\footnote{permanent address: Centro de
F\'{\i}sica, Instituto Venezolano de Investigaciones Cient\'{\i}ficas IVIC,
Apdo. 21827, Caracas 1020-A Venezuela\\ e-mail
address: villalba@cpt.univ-mrs.fr, villalba@dino.conicit.ve}

\end{center}
\vspace{1.5 cm}

\centerline{\bf Abstract}

\vspace{0.3cm}

The Dirac equation in spherically symmetric fields is separated in two
different tetrad frames. One is the standard cartesian (fixed) frame
and the second one is the diagonal (rotating) frame. After
separating variables in the Dirac equation in spherical coordinates, and
solving the corresponding eingenvalues equations associated with the angular
operators, we obtain that the spinor solution in the rotating frame can be
expressed in terms of Jacobi polynomials, and it is related to the standard
spherical harmonics, which are the basis solution of the angular momentum
in the Cartesian tetrad, by a similarity transformation

\vskip 1truecm

\noindent Key-Words : 3.65, 11.10, 4.90.

\bigskip

\noindent April 1994

\noindent CPT-94/P.3027

\noindent anonymous ftp or gopher: cpt.univ-mrs.fr

\end{titlepage}

\section{Introduction}

\setcounter{equation}{0}

The Dirac equation is a system of four coupled partial differential equation
which describes the relativistic electron and other spin 1/2 particles.
Despite the remarkable effort made during the last decades in order to find
exact solutions for the relativistic electron in the presence of external
fields, the amount of solvable problems is relatively scarce, being the
Coulomb problem \cite{Coulomb}, the plane wave \cite{volkov}, and some
electromagnetic configurations \cite{Bagrov} the most representative
examples. The problem of finding exact solutions of the Dirac equation is
closely related to the possibility of accomplishing a complete separation of
variables of the Dirac equation.

During the last decades there have been different approaches to tackling the
problem of separating variables in the Dirac equation in the presence of
external fields; we have the St\"ackel spaces method developed by Bagrov and
co-workers (\cite{Bagrov} based on the experienced acquired in separating
variables in the Klein Gordon equation), the theory of R-separability by
Miller \cite{Miller1,Miller2}, and matrix modification of the St\"ackel
proposed by Cook \cite{Cook} Recently an algebraic method of separation of
variables \cite{alg1,alg2}has been successfully applied in different
gravitational and electromagnetic configurations. The main idea of the
algebraic method of separation is to try to reduce the original Dirac
equation, which in the presence of electromagnetic fields takes the form

\begin{equation}
\label{Dirac}\{D\}\Psi =\left\{ {\tilde \gamma }^\mu (\partial _\mu -\Gamma
_\mu -iA_\mu )+{\rm m}\right\} \Psi ~=~0
\end{equation}
to a new equivalent equation according the following scheme

\begin{equation}
\label{equi}\{D\}\Psi =0\Rightarrow {\cal F}\{D\}\Gamma \Gamma ^{-1}\Psi
=(\hat K_1+\hat K_{2)}\Phi =0
\end{equation}
with

\begin{equation}
\Psi =\Gamma \Phi
\end{equation}
where \^K$_1$ and \^K$_2$ are first-order commuting matrix differential
operators

\begin{equation}
\label{conm}\left[ \hat K_1,\hat K_2\right] _{-}=0
\end{equation}
depending each one on different space-time variables, $\Gamma $ is a
constant nonsingular matrix, and ${\cal F}$ that we have to find (if they
exist) in order to fulfill the condition (\ref{conm}). Depending on the
problem to solve there are two different schemes of separation of variables
(1) Consecutive separation of variables when one variable is separated from
the remaining three variables, then one variable is separated from the other
two variables, and finally, the remaining two variables are separated. (2)
Pairwise separation of variables, when two space-time variables are
separated from the another two ones, and the resulting operators are also
separated. In the present article we focus our attention to the Dirac
equation in the presence of a central field, like the Hydrogen Atom where a
complete separation of variables is possible. In this case, using a pairwise
scheme of separation, the angular dependence is completely separated from
the other two variables.
The article is structured as follows. In Sec. II we separate variables in
the Dirac equation. In Sec III. we solve the angular equation in the local
rotating frame in terms of Jacobi Polynomials In Sec. IV we establish the
relation between the spinor solution in the diagonal (rotating) tetrad gauge
and the fixed Cartesian one.

\section{Separation of variables}

\setcounter{equation}{0}

In this section we proceed to separate variables in the Dirac equation (\ref
{Dirac}), where $\tilde \gamma ^\mu $ are the curved gamma matrices
satisfying the relation,

\begin{equation}
\label{10}\left\{ {\tilde \gamma }^\mu ,\tilde \gamma ^\nu \right\}
_{+}=2g^{\mu \nu }
\end{equation}

\noindent and $\Gamma _\mu $ are the spin connections \cite{Brill}.

\begin{equation}
\label{11}\Gamma _\lambda ={\frac 14}g_{\mu \alpha }[(\partial b_\nu ^\beta
/\partial x^\lambda )a_\beta ^\alpha -\Gamma _{\nu \lambda }^\alpha ]s^{\mu
\nu }
\end{equation}
with

\begin{equation}
\label{12}s^{\mu \nu } = {\frac{1}{2}}(\tilde\gamma ^{\mu } \tilde\gamma
^{\nu } -\tilde\gamma ^{\nu } \tilde\gamma ^{\mu })
\end{equation}

\noindent and the matrices $b^{\beta }_{\alpha }$ , $a^{\alpha }_{\beta }$
establish the connection between the Dirac matrices $(\tilde\gamma )$ on a
curved spacetime and the Minkowski space $(\gamma)$ Dirac matrices as
follows:

\begin{equation}
\label{13}\tilde \gamma _\mu =b_\mu ^\alpha {\gamma }_\alpha \qquad \tilde
\gamma ^\mu =a_\beta ^\mu {\gamma }^\beta
\end{equation}
If we choose to work in the fixed Cartesian gauge, then spinor connections
are zero and the $\tilde \gamma $ matrices take the form
$$
\tilde \gamma ^0={\gamma }^0=\bar \gamma ^0,\ \tilde \gamma ^1=\left[
(\gamma ^1\cos \varphi +\gamma ^2\sin \varphi )\sin \vartheta +\gamma ^3\cos
\vartheta \right] =\bar \gamma ^1,
$$
\begin{equation}
\label{carte}\tilde \gamma ^2=\frac 1r\left[ (\gamma ^1\cos \varphi +\gamma
^2\sin \varphi )\cos \vartheta -\gamma ^3\sin \vartheta \right] =\frac{\bar
\gamma ^2}r,\quad
\end{equation}
$$
\tilde \gamma ^3=\frac 1{r\sin \vartheta }(-\gamma ^1\sin \varphi +\gamma
^2\cos \varphi )=\frac{\bar \gamma ^3}{r\sin \vartheta }
$$
and the Dirac equation in the fixed tetrad frame (\ref{carte}) takes the
form
\begin{equation}
\left\{ \bar \gamma ^0(\partial _t+iV(r)+\bar \gamma ^1\partial _r+\frac{%
\bar \gamma ^2}r\partial _\vartheta +\frac{\bar \gamma ^3}{r\sin \vartheta }%
\partial _\varphi +{\rm m}\right\} \tilde \Psi ~=~0
\end{equation}
In order to separate variables in the Dirac equation, we are going to work
in the diagonal tetrad gauge where the gamma matrices $\tilde \gamma _d$
take the form

\begin{equation}
\label{diag}\tilde \gamma _d^0=\gamma ^0,\ \tilde \gamma _d^1=\gamma ^1,\
\tilde \gamma _d^2=\frac 1r\gamma ^2,\ \tilde \gamma _d^3=\frac 1{r\sin
\vartheta }\gamma ^3
\end{equation}
Since the curvilinear matrices $\tilde \gamma ^\mu $ matrices and $\tilde
\gamma _d$ satisfy the same anticommutation relations, they are related by a
similarity transformation, unique up to a factor. In the present case we
choose this factor in order to eliminate the spin connections in the
resulting Dirac equation. The transformation S can be written as \cite
{victor}

\begin{equation}
\label{vest}S=\frac 1{r(\sin \vartheta )^{1/2}}\exp (-\frac \varphi 2\gamma
^1\gamma ^2)\exp (-\frac \vartheta 2\gamma ^3\gamma ^1){\sf a=S}_0{\sf a}
\end{equation}
where ${\sf a}$ is the constant non singular matrix given by the expression
\begin{equation}
{\sf a}=\frac 12(\gamma ^1\gamma ^2-\gamma ^1\gamma ^3+\gamma ^2\gamma ^3+I)
\end{equation}
which applied on the gamma's acts as follows

\begin{equation}
\label{a}{\sf a}\gamma ^1{\sf a}^{-1}=\gamma ^3,\ {\sf a}\gamma ^2{\sf a}%
^{-1}=\gamma ^1,\ {\sf a}\gamma ^3{\sf a}^{-1}=\gamma ^2,
\end{equation}
the transformation S acts on the curvilinear $\tilde \gamma $ matrices,
reducing them to the rotating diagonal gauge as follows

\begin{equation}
S^{-1}\tilde \gamma ^\mu S=g^{\mu \mu }\gamma ^\mu =\tilde \gamma _d^\mu
\quad {\rm (no\ summation)}
\end{equation}
then, the Dirac equation in spherical coordinates, with the radial potential
$V(r)$, in the local rotating frame reads

\begin{equation}
\label{uno}\left\{ \gamma ^0\partial _t+\gamma ^1\partial _r++\frac{\gamma ^2%
}r\partial _\vartheta +\frac{\gamma ^3}{r\sin \vartheta }\partial _\varphi +%
{\rm m}+i\gamma ^0V(r)\right\} \Psi =0
\end{equation}

\noindent where we have introduced the spinor $\Psi $, related to $\tilde
\Psi$ by the expression

\begin{equation}
\label{16}\tilde \Psi =S\Psi =S_0{\sf a}\Psi
\end{equation}

\noindent and $\gamma ^\mu $ are the standard Dirac flat matrices.
\noindent Applying the algebraic method of separation of variables \cite
{alg1,alg2}, it is possible to write eq. (\ref{uno}) as a sum of two first
order linear differential operators $\hat K_1$ , $\hat K_2$ satisfying the
relation

\begin{equation}
\label{17}\left[ \hat K_1,\hat K_2\right] =0,\hspace{2cm}\left\{ \hat K_1+{%
\hat K}_2\right\} \Phi =0
\end{equation}

\begin{equation}
\label{18}{\hat K}_1\Phi =\lambda \Phi =-{\hat K}_2\Phi
\end{equation}
then, if we separate the time and radial dependence from the angular one, we
obtain

\begin{equation}
\label{dos}{\hat K}_2\Phi =\left[ \gamma ^2\partial _\vartheta +\frac{\gamma
^3}{\sin \vartheta }\partial _\varphi \right] \gamma ^0\gamma ^1\Phi
=-i\kappa \Phi
\end{equation}

\begin{equation}
\label{quatre}{\hat K}_1\Phi =r\left[ \gamma ^0\partial _t+\gamma ^1\partial
_r+m+i\gamma ^0V\right] \gamma ^0\gamma ^1\Phi =i\kappa \Phi
\end{equation}
with

\begin{equation}
\label{ici}\Psi =\gamma ^0\gamma ^1\Phi
\end{equation}
where we have made the identification $i\kappa =\lambda .$ Notice that (\ref
{dos}) is the angular momentum \^K obtained by Brill and Wheeler \cite{Brill}.
Here it is necessary to remark that the operator ${\hat K}_2$ appearing in (%
\ref{dos}) is not singled valued and does not satisfy the properties of a
''good'' angular momentum operator. The true angular operator should be
obtained from ${\hat K}_2$ with the help of the transformation S (\ref{vest}%
)
\begin{equation}
S\left[ \gamma ^2\partial _\vartheta +\frac{\gamma ^3}{\sin \vartheta }%
\partial _\varphi \right] \gamma ^0\gamma ^1S^{-1}S\Phi =-i\kappa S\Phi
\end{equation}
since S$\gamma ^\mu S^{-1}=\bar \gamma ^\mu $, we obtain

\begin{equation}
\label{zero}\left[ \bar \gamma ^2\bar \gamma ^0\bar \gamma ^1\partial
_\vartheta +\frac{\bar \gamma ^3\bar \gamma ^0\bar \gamma ^1}{\sin \vartheta
}\partial _\varphi +\bar \gamma ^2\bar \gamma ^0\bar \gamma ^1S\partial
_\vartheta S^{-1}+\frac{\bar \gamma ^3\bar \gamma ^0\bar \gamma ^1}{\sin
\vartheta }S\partial _\varphi S^{-1}\right] \tilde \Phi =-i\kappa \tilde
\Phi
\end{equation}
where S$\Phi =\tilde \Phi $
using the explicit form of the transformation S given by (\ref{vest}) we
have

\begin{equation}
\label{pri}S\partial _\vartheta S^{-1}=\frac 12\cot \vartheta +\frac 12(\cos
\varphi -\gamma ^1\gamma ^2\sin \varphi )\gamma ^3\gamma ^1
\end{equation}

\begin{equation}
\label{se}S\partial _\varphi S^{-1}=\frac 12\gamma ^1\gamma ^2
\end{equation}
substituting (\ref{pri}) and (\ref{se}) into (\ref{zero})

\begin{equation}
\label{mom}\left[ \bar \gamma ^0\bar \gamma ^2\bar \gamma ^1\partial
_\vartheta +\frac{\bar \gamma ^0\bar \gamma ^3\bar \gamma ^1}{\sin \vartheta
}\partial _\varphi +\bar \gamma ^0\right] \tilde \Phi =i\kappa \tilde \Phi
\end{equation}
expression that can be written as follows

\begin{equation}
\label{ca}\hat K=\gamma ^0\left[ ({\bf \sigma \hat L)+}1\right]
\end{equation}
The operator $\hat K$ satisfies the relation

\begin{equation}
\hat K^2=\hat J^2+\frac 14,\ \hat J^2=(L+\frac 12{\bf \sigma )}^2
\end{equation}
and the eigenvalue $k$ is related to $j$ as follows

\begin{equation}
k^2=j(j+1)+\frac 12=(j+\frac 12)^2
\end{equation}
Now we proceed to decouple the equation (\ref{dos}) governing the angular
dependence of the Dirac spinor In order to simplify the resulting equations,
we choose to work with the auxiliary spinor $\bar \Phi $ related to $\Phi $
as follows

\begin{equation}
\bar \Phi ={\sf a}\Phi
\end{equation}
then the equation (\ref{dos}) takes the form

\begin{equation}
\label{cinq}\left[ \gamma ^1\partial _\vartheta +\frac{\gamma ^2}{\sin
\vartheta }\partial _\varphi \right] \gamma ^0\gamma ^3\bar \Phi =-i\kappa
\bar \Phi
\end{equation}
In order to reduce the equation (\ref{cinq}) to a system of ordinary
differential equations, we choose to work in the following representation
for the gamma matrices, \cite{Jauch}

\begin{equation}
\label{tres}\gamma ^i=\left(
\begin{array}{cc}
0 & \sigma ^i \\
\sigma ^i & 0
\end{array}
\right) ,\gamma ^0=\left(
\begin{array}{cc}
-i & 0 \\
0 & i
\end{array}
\right) ,
\end{equation}
Then, substituting (\ref{tres}) into (\ref{cinq}) we obtain,
\begin{equation}
\label{ang1}\left[ \sigma ^2\partial _\vartheta -i\frac{\sigma ^1}{\sin
\vartheta }m\right] \bar \Phi _1=-ik\bar \Phi _1
\end{equation}

\begin{equation}
\label{ang2}\left[ -\sigma ^2\partial _\vartheta +i\frac{\sigma ^1}{\sin
\vartheta }m\right] \bar \Phi _2=-ik\bar \Phi _2
\end{equation}
with

\begin{equation}
\bar \Phi =\left(
\begin{array}{c}
\bar \Phi _1 \\
\bar \Phi _2
\end{array}
\right)
\end{equation}

\section{Solution of the angular equations}

\setcounter{equation}{0}

In this section we are going to solve the systems of equations (\ref{ang1})
and (\ref{ang2}). It is not difficult to see that the spinor $\bar \Phi _2$
is proportional to $\sigma ^3\bar \Phi _1$, this property allows us to
consider only the system (\ref{ang1}). Using the standard Pauli matrices ,
we have that (\ref{ang1}) reduces to

\begin{equation}
\label{tsi1}k\sin \vartheta \xi _1-m\xi _2-\sin \vartheta \frac{d\xi _2}{%
d\vartheta }=0
\end{equation}

\begin{equation}
\label{tsi2}k\sin \vartheta \xi _2-m\xi _1+\sin \vartheta \frac{d\xi _1}{%
d\vartheta }=0
\end{equation}
with

\begin{equation}
\label{Fi1}\bar \Phi _1=\left(
\begin{array}{c}
\xi _1 \\
\xi _2
\end{array}
\right)
\end{equation}
In order to solve the coupled system of equations (\ref{tsi1})-(\ref{tsi2})
we make the following ansatz,

\begin{equation}
\label{3'}\quad \xi _1=i(\sin \frac \vartheta 2)^m(\cos \frac \vartheta
2)^{m+1}q(\vartheta ),\ \xi _2=(\sin \frac \vartheta 2)^{m+1}(\cos \frac
\vartheta 2)^mf(\vartheta )
\end{equation}
Substituting (\ref{3'}) into (\ref{tsi1})
and (\ref{tsi2}) we obtain,

\begin{equation}
\label{6'}ikq(x)-(m+\frac 12)f(x)+(1-x)\frac{df(x)}{dx}=0
\end{equation}

\begin{equation}
\label{7'}ikf(x)+(m+\frac 12)q(x)+(1+x)\frac{dq(x)}{dx}=0
\end{equation}
where we have made the change of variable $x=\cos \vartheta ,$
The normalizable solution of the system (\ref{6'})-(\ref{7'}) can be
expressed in terms of Jacobi Polynomials P$_k^{(\alpha ,\beta )}$\cite
{Magnus}

\begin{equation}
\label{J2}\xi _1=c_0(\sin \frac \vartheta 2)^m(\cos \frac \vartheta
2)^{m+1}P_n^{(m-1/2,m+1/2)}
\end{equation}

\begin{equation}
\label{J1}\xi _2=c_0(\sin \frac \vartheta 2)^{m+1}(\cos \frac \vartheta
2)^mP_n^{(m+1/2,m-1/2)}
\end{equation}
where c$_0$ is an arbitrary constant, which can be fixed in order to
normalize the angular dependence of the spinor $\Phi $ according to the
product

\begin{equation}
\label{int}2\pi \int_0^{2\pi }\Phi _k^{\mu \dagger }\Phi _{k^{\prime }}^\mu
\ d\vartheta =\delta _{kk^{\prime }}
\end{equation}
and $n$ is given by the expression

\begin{equation}
\label{ene}n=-m-\frac 12+k,\quad n=0,1,2....
\end{equation}
then, using the orthogonality relation for the Jacobi polynomials \cite
{Magnus}

\begin{eqnarray}
\int^{1}_{-1} (1-x)^\alpha (1+x)^\beta P_n^{(\alpha ,\beta )}(x)P_m^{(\alpha
,\beta
)}(x)dx= \nonumber  \\
\frac{2^{\alpha +\beta +1}}{2n+\alpha +\beta +1}\frac{\Gamma
(n+\alpha +1)\Gamma (n+\beta +1)}{\Gamma (n+1)\Gamma (n+\alpha +\beta +1)}%
\delta _{mn}
\end{eqnarray}
we obtain

\begin{equation}
\label{co}c_0=\frac{\sqrt{n!(n+2m)}!}{(n+m-1/2)!\sqrt{2\pi }}
\end{equation}
Now, we are going to solve the system of equation (\ref{quatre}), governing
the radial dependence of the spinor $\Psi $ solution of the Dirac equation.
This equation can be written in the form,

\begin{equation}
\left( -\gamma ^3\partial _t+\gamma ^0\partial _r+m\gamma ^0\gamma
^3-i\gamma ^3V(r)+i\frac \kappa r\right) \bar \Phi =0
\end{equation}

Using the representation for the gamma matrices given by ($\ref{tres}$), we
obtain the following system of equations

\begin{equation}
\label{lap}\left( d_r+\frac kr\right) \bar \Phi _1-\sigma ^3(E-V(r)-m)\bar
\Phi _2=0
\end{equation}

\begin{equation}
\label{las}\left( -d_r+\frac kr\right) \bar \Phi _2-\sigma ^3(E-V(r)+m)\bar
\Phi _1=0
\end{equation}
therefore, the general form of the spinor $\bar \Phi $ is given by the
expression

\begin{equation}
\label{sol}\bar \Phi =c_0(\sin \frac \vartheta 2)^m(\cos \frac \vartheta
2)^me^{i(m\varphi -Et)}\left(
\begin{array}{c}
c(\cos \frac \vartheta 2)P_n^{(m-1/2,m+1/2)}A(r) \\
c(\sin \frac \vartheta 2)P_n^{(m+1/2,m-1/2)}A(r) \\
(\cos \frac \vartheta 2)P_n^{(m-1/2,m+1/2)}B(r) \\
-(\sin \frac \vartheta 2)P_n^{(m+1/2,m-1/2)}B(r)
\end{array}
\right)
\end{equation}
where c$_0$ is given by (\ref{co}), $c$ is a constant, and $A(r),B(r)$
satisfy the system of equations

\begin{equation}
\left( d_r+\frac kr\right) A(r)-(E-V(r)-m)B(r)=0
\end{equation}

\begin{equation}
\left( -d_r+\frac kr\right) B(r)-(E-V(r)+m)A(r)=0
\end{equation}
obviously, the form A(r) and B(r) will depend on the potential V(r) . Among
the solvable problems we have the Coulomb case where the radial dependence
can be expressed in terms of confluent hypergeometric functions.\cite
{Davydov}

Regarding the eigenvalues $m$ of the projection of the angular momentum
operator $-i\partial _\varphi $ we have that since

\begin{equation}
S_z(\varphi +\pi )=-S_z(\varphi )
\end{equation}
and the spinor $\tilde \Psi $ is single valued, then $\Psi (\varphi +2\pi
)=-\Psi (\varphi ),$ and therefore $m$ takes half integer values

\begin{equation}
m=N+\frac 12,\ N=0,\ \pm 1,\ \pm 2...
\end{equation}

\section{Transformation of the spinor}

\setcounter{equation}{0}

In this section we proceed to show that the angular dependence of the spinor
solution of the Dirac equation (\ref{uno}), which is given by the expression
(\ref{sol}), reduces to the standard spherical spinors when we apply the
transformation S given by (\ref{vest})
Since the spinor $\tilde \Psi $, expressed in the Cartesian tetrad gauge, is
related to the spinor solution $\Psi $ in the local rotating frame by means
of (\ref{16}), we have

\begin{equation}
\label{s1}S_0\gamma ^0\gamma ^3\bar \Phi =\tilde \Psi
\end{equation}
where we have used (\ref{ici}) and (\ref{a}). Using the representation (\ref
{tres}) and the spinor (\ref{sol}), we obtain

\begin{equation}
\label{s2}\tilde \Psi =\left(
\begin{array}{c}
\tilde \Psi ^{+} \\
\tilde \Psi ^{-}
\end{array}
\right) -i\frac{c_0}{\sqrt{2}r}(\sin \frac \vartheta 2\cos \frac \vartheta
2)^{m-1/2}S_\vartheta (\vartheta )e^{im\varphi }\left(
\begin{array}{c}
(\cos \frac \vartheta 2)P_n^{(m-1/2,m+1/2)}B(r) \\
(\sin \frac \vartheta 2)P_n^{(m+1/2,m-1/2)}B(r) \\
-c(\cos \frac \vartheta 2)P_n^{(m-1/2,m+1/2)}A(r) \\
c(\sin \frac \vartheta 2)P_n^{(m+1/2,m-1/2)}A(r)
\end{array}
\right)
\end{equation}

\begin{equation}
\label{s3}S_\varphi (\varphi )S_\vartheta (\vartheta )=\left(
\begin{array}{cccc}
e^{-i\varphi /2} & 0 & 0 & 0 \\
0 & e^{i\varphi /2} & 0 & 0 \\
0 & 0 & e^{-i\varphi /2} & 0 \\
0 & 0 & 0 & e^{i\varphi /2}
\end{array}
\right) \left(
\begin{array}{cccc}
\cos \frac \vartheta 2 & \sin \frac \vartheta 2 & 0 & 0 \\
-\sin \frac \vartheta 2 & \cos \frac \vartheta 2 & 0 & 0 \\
0 & 0 & \cos \frac \vartheta 2 & \sin \frac \vartheta 2 \\
0 & 0 & -\sin \frac \vartheta 2 & \cos \frac \vartheta 2
\end{array}
\right)
\end{equation}
substituting (\ref{s3}) into (\ref{s2}) and taking into account the relations

\begin{equation}
\label{s4}(1-x)P_n^{(\alpha +1,\beta )}(x)+(1+x)P_n^{(\alpha ,\beta
+1)}(x)=2P_n^{(\alpha ,\beta )}(x)
\end{equation}
\begin{equation}
\label{s5}P_n^{(\alpha ,\beta -1)}(x)=P_n^{(\alpha -1,\beta
)}(x)+P_{n+1}^{(\alpha ,\beta )}(x)
\end{equation}
we find that the upper two components of the spinor $\tilde \Psi $ (\ref{s2}%
) take the form

\begin{equation}
\label{s6}\tilde \Psi ^{+}=-i\frac{c_0}{\sqrt{2}r}(\sin \frac \vartheta
2\cos \frac \vartheta 2)^{1/2}\left(
\begin{array}{c}
(\sin \frac \vartheta 2\cos \frac \vartheta
2)^{m-1/2}P_n^{(m-1/2,m-1/2)}(\cos \vartheta )e^{i(m-1/2)\varphi }B(r) \\
(\sin \frac \vartheta 2\cos \frac \vartheta
2)^{m+1/2}P_{n-1}^{(m+1/2,m+1/2)}(\cos \vartheta )e^{i(m+1/2)\varphi }B(r)
\end{array}
\right)
\end{equation}
with the help of the relation

\begin{equation}
\label{s7}Y_{l.m}(\vartheta ,\varphi )=\frac 1{2^ml!}\sqrt{\frac{2l+1}{4\pi }%
(l-m)!(l+m)!}(1-x^2)^{m/2}P_{l-m}^{(m,m)}(x)e^{im\varphi }
\end{equation}
and

\begin{equation}
\label{s8}P_l^{-m}(x)=(-1)^m\frac{(l-m)!}{(l+m)!}P_l^m(x)
\end{equation}
we find that $\tilde \Psi ^{+}$ can be expressed in terms of spherical
harmonics Y$_{k,m}(\vartheta ,\varphi )$ as follows

\begin{equation}
\label{s9}\tilde \Psi ^{+}=-i\frac 1r(2k+1)^{-1/2}\left(
\begin{array}{c}
(k+m+1/2)^{1/2}Y_k^{m-1/2}(\vartheta ,\varphi )B(r) \\
(k-m+1/2)Y_k^{m+1/2}(\vartheta ,\varphi )B(r)
\end{array}
\right)
\end{equation}
since the transformation S$_0$ acting on the spinor $\Psi $ has a block
form, it is possible to write it as follows

\begin{equation}
\label{s10}S_0=\left(
\begin{array}{cc}
\hat S_0 &  \\
& \hat S_0
\end{array}
\right) ,\ \hat S_0=\frac 1{r(\sin \vartheta )^{1/2}}\left(
\begin{array}{cc}
e^{-i\varphi /2} & 0 \\
0 & e^{i\varphi /2}
\end{array}
\right) \left(
\begin{array}{cc}
\cos \frac \vartheta 2 & \sin \frac \vartheta 2 \\
-\sin \frac \vartheta 2 & \cos \frac \vartheta 2
\end{array}
\right)
\end{equation}
then, if we set c=-1, we have

\begin{equation}
\label{s11}\tilde \Psi ^{-}=\hat S_0\bar \Phi ^{-}=\hat S_0\sigma _3\bar
\Phi ^{+}=\hat S_0\sigma ^3\hat S_0^{-1}\hat S_0\bar \Phi ^{+}={\bf \sigma
.\hat e}_r\tilde \Psi ^{+}
\end{equation}
and the lower components $\tilde \Psi ^{-}$ of the spinor $\tilde \Psi $
read

\begin{equation}
\label{s12}\tilde \Psi ^{-}=i\frac 1r(2k+1)^{-1/2}\left(
\begin{array}{c}
(k-m+1/2)^{1/2}Y_k^{m-1/2}(\vartheta ,\varphi )A(r) \\
-(k+m+1/2)Y_k^{m+1/2}(\vartheta ,\varphi )A(r)
\end{array}
\right)
\end{equation}
the results (\ref{s9}) and (\ref{s12}) are in terms of the well known
spherical spinors as in almost all the textbooks on relativistic quantum
mechanics. The basis solution in terms of Jacobi Polynomials has been of
utility in analyzing the Dirac spinor in the presence of a magnetic monopole,
also it can be of help in discussing non-central
interactions.

\vspace{2cm}

\centerline{\bf acknowledgments}

\medskip

\noindent The author wishes to express his indebtedness to the Centre de
Physique
Th\'eorique for the suitable conditions of work. Also the author wishes to
acknowledge to the CONICIT of Venezuela and to the Fundaci\'on Polar for
financial support.

\end{document}